\documentclass[12pt]{article}
\usepackage{psfig}
\usepackage{epsfig}
\usepackage{rotating}

\newcommand{\epsla}{\epsilon\hspace{-0.45em}/}
\newcommand{\qsla}{q\hspace{-0.45em}/}

\def\ba{\begin{eqnarray}}
\def\ea{\end{eqnarray}}
\def\dd{{\rm d}}
\def\vep{\epsilon}
\def\ep{\epsilon}
\def\order#1{{\cal O}\left(#1\right)}

\newcommand{\bea}{\begin{eqnarray}}
\newcommand{\eea}{\end{eqnarray}}
\newcommand{\bd}{\begin{displaymath}}
\newcommand{\ed}{\end{displaymath}}
\newcommand{\be}{\begin{equation}}
\newcommand{\ee}{\end{equation}}
\newcommand{\ord}{{\cal O}}

\def\bbuildrel#1_#2^#3{\mathrel{\mathop{\kern 0pt#1}\limits_{#2}^{#3}}}
\def\slash#1{\setbox0=\hbox{$#1$}#1\hskip-\wd0\dimen0=5pt\advance
       \dimen0 by-\ht0\advance\dimen0 by\dp0\lower0.5\dimen0\hbox
         to\wd0{\hss\sl/\/\hss}}

\newcommand{\f}{\frac}

\newcommand{\beq}{\begin{equation}}
\newcommand{\eeq}{\end{equation}}

\def\con{\ifmmode \hbox{\bf*} \else{\bf*}\fi}   
\def\scon{\ifmmode \hbox{\footnotesize\rm\bf*} \else{\footnotesize\rm\bf*}\fi}

\def\0#1{\relax\ifmmode\mathaccent"7017{#1}
        \else\accent23#1\relax\fi}              

\begin{document}


\author{ {\large\bf Andrzej J.~Buras${}^{1}$, Andrzej Czarnecki${}^{2}$},
    \\
  {\large\bf Miko{\l}aj Misiak${}^{3}$
    and J\"org Urban${}^{1}$} \\
  \ \\
  {\small ${}^{1}$ Physik Department,
    Technische Universit{\"a}t M{\"u}nchen,} \\
  {\small D-85748 Garching, Germany} \\
  {\small ${}^{2}$ Physics Department,
    University of Alberta,} \\
  {\small Edmonton, Alberta, Canada T6G 2J1} \\
    {\small ${}^{3}$ Institute of Theoretical Physics, Warsaw University
    } \\
  {\small Ho\.za 69, 00-681 Warsaw, Poland}
}
\date{May 16, 2001}
\title{
{\normalsize\sf
\rightline {hep-ph/0105160}
\rightline{TUM-HEP-415/01}
\rightline{Alberta Thy 10-01}
\rightline{IFT-15/2001}
}
\bigskip
\date{\today}
{\LARGE\bf
\bigskip
Two--Loop Matrix Element of the Current--Current Operator in the Decay
\boldmath{$B\to X_s\gamma$} 
}}

\maketitle
\thispagestyle{empty}

\phantom{xxx} \vspace{-9mm}

\begin{abstract}

We evaluate the important two--loop matrix element $\langle s
\gamma|Q_2|b\rangle$ of the operator $\left(\bar c \gamma^\mu P_L
b\right) \left(\bar s \gamma_\mu P_L b\right)$ contributing to the
inclusive radiative decay $B\rightarrow X_s \gamma$. The calculation
is performed in the NDR
scheme, by means of asymptotic expansions method. The result is given
as a series in $z\equiv m_c^2/m_b^2$ up to ${\cal O}(z^6)$. 
We confirm the result of
Greub, Hurth and Wyler obtained by a different method up to ${\cal
  O}(z^3)$.
Higher--order terms are found to be numerically insignificant.

\end{abstract}

\newpage
\setcounter{page}{1}
\setcounter{footnote}{0}  

\section{Introduction}
\setcounter{equation}{0} 

The radiative decay $B\to X_s\gamma$ plays an important role in the
present tests of the Standard Model (SM) and of its extensions
\cite{Campbell:1982rg}.  In particular, in the supersymmetric
extensions of the SM, the best bounds on several new parameters come
from the data on $B\to X_s\gamma$ \cite{ENO01}.

The short distance QCD effects are very important for this decay.
They are known to enhance the branching ratio ${\rm Br}(B\to
X_s\gamma)$ by roughly a factor of three, as first pointed out in
\cite{Bert,Desh}. Since these first analyses, a lot of progress has
been made in calculating these important QCD effects in the
renormalization group improved perturbation theory, beginning with the
work in \cite{Grin,Odon}.  Let us briefly summarize this progress.

A peculiar feature of the renormalization group analysis in $B\to
X_s\gamma$ is that the mixing under infinite renormalization between
the four-fermion operators $(Q_1,\ldots,Q_6)$ and the magnetic penguin
operators $(Q_{7\gamma},Q_{8G})$, which govern this decay, vanishes at
the one-loop level. Consequently, in order to calculate the
coefficients $C_{7\gamma}(\mu_b)$ and $C_{8G}(\mu_b)$ at
$\mu_b={\cal O}(m_b)$ in the leading
logarithmic approximation, two-loop calculations of ${\cal{O}}(e
g^2_s)$ and ${\cal{O}}(g^3_s)$ are necessary. Such calculations were
completed in \cite{CFMRS:93,CFRS:94} and confirmed in
\cite{CCRV:94a,CCRV:94b,CZMM}. Earlier analyses contained either
additional approximations or mistakes.

It turns out that the leading order expression for the branching ratio
${\rm Br}(B\to X_s\gamma)$ suffers from sizable renormalization scale
uncertainties \cite{AG1,BMMP:94} implying that a complete NLO analysis
including also dominant higher order electroweak effects to this decay
is mandatory.  By 1998, the main ingredients of such an analysis had
been calculated.  It was a joint effort of many groups:
\begin{itemize}
\item
The $\ord(\alpha_s)$ corrections to $C_{7\gamma}(\mu_W)$ and
$C_{8G}(\mu_W)$ were first calculated in \cite{Yao1} and subsequently
confirmed by several groups \cite{GH97}--\cite{BMU99}.
\item
The two-loop mixing involving the four fermion operators $Q_1,\ldots,Q_6$
and the two-loop mixing in the sector $(Q_{7\gamma},Q_{8G})$ was
calculated in \cite{ACMP}--\cite{ROMA2} and \cite{MisMu:94},
respectively.  The very difficult three-loop mixing between the set
$(Q_1,\ldots,Q_6)$ and the operators $(Q_{7\gamma},Q_{8G})$ was calculated
in \cite{CZMM}.
\item
Leading order matrix elements $\langle s\gamma \;{\rm gluon}|Q_i|
b\rangle$ were calculated in \cite{AG2,Pott}, and the challenging
two-loop calculation of $\langle s\gamma |Q_2| b\rangle$ was presented
in \cite{GREUB}.
\item
Higher order electroweak corrections were incorporated with increasing
level of sophistication in \cite{CZMA}--\cite{GH00}.
\end{itemize}
In addition, non-perturbative corrections were calculated in
\cite{FLS94}--\cite{BIR98}. The most recent analysis of $B\to
X_s\gamma$ incorporating all these calculations can be found in
\cite{GAMMIS}.

Now, among the perturbative ingredients listed above, three have been
calculated only by one group. These are
\begin{itemize}
\item
The two-loop mixing
in the sector $(Q_{7\gamma},Q_{8G})$
\cite{MisMu:94}.
\item
The three-loop mixing between
the set $(Q_1,\ldots,Q_6)$ and the operators $(Q_{7\gamma},Q_{8G})$
 \cite{CZMM}.
\item
The two-loop matrix element 
$\langle s\gamma |Q_2| b\rangle$ \cite{GREUB}.
\end{itemize}
Moreover, 
\begin{itemize}
\item
The two-loop matrix elements of the QCD penguin operators $\langle
s\gamma|Q_i|b\rangle$ with \linebreak $i=3,\ldots,6$ have not yet been
calculated.
\end{itemize}

It should be emphasized that all these four ingredients enter not only
the analysis of $B\to X_s\gamma$ in the SM but are also necessary
ingredients of any analysis of this decay in the extensions of this
model.  It is therefore desirable to check the first three
calculations and to perform the last one.

In the present paper, we will make the first step in this direction by
calculating the two-loop matrix element $\langle s\gamma |Q_2|
b\rangle$ using the method of asymptotic expansions. This matrix
element turned out \cite{{GREUB},{GAMMIS}} to be the most important
ingredient of the NLO--analysis for Br$(B\rightarrow X_s\gamma)$
enhancing this branching ratio by roughly $20\%$. In \cite{GREUB}, the
matrix element $\langle s\gamma |Q_2| b\rangle$ was found by
applying the Mellin--Barnes representation to certain internal
propagators, and the result was  presented as an expansion in
$z=m_c^2/m_b^2$ up to and including terms $\ord(z^3)$. In order to be
sure about the convergence of this expansion, we will include also the
terms $\ord(z^4)$, $\ord(z^5)$ and $\ord(z^6)$.


\section{Two-Loop Contribution to $\langle s\gamma|Q_2|b\rangle$.}
\subsection{Preface}

The effective Hamiltonian for the process $b\rightarrow s\gamma$ is
given by
\begin{equation}
\label{effham}
H_{\rm eff} (b\rightarrow s\gamma) = -\frac{4 G_F}{\sqrt{2}} V_{tb}
V_{ts}^* \sum_{i=1}^8 C_i(\mu) Q_i.
\end{equation}
Here $G_F$ is the Fermi constant, $V_{ij}$ are the CKM matrix elements
and $C_i$ are the Wilson coefficients of the operators $Q_i$ evaluated
at $\mu={\cal O}(m_b)$. We have dropped the negligible contributions
proportional to $V_{ub} V_{us}^*$. The full list of operators in
(\ref{effham}) can be found in \cite{GREUB}. In the present paper we
need only the expressions for two of them. These are 
\begin{eqnarray}
Q_2 &=& \left(\bar c \gamma^\mu P_L b\right) 
        \left(\bar s \gamma_\mu P_L b\right),\\
Q_7 &=& \frac{e}{16 \pi^2} \bar s_\alpha \sigma^{\mu \nu} 
        (m_b (\mu) P_R + m_s (\mu) P_L) b_\alpha F_{\mu \nu},
\end{eqnarray}
where $P_L=(1-\gamma_5)/2$, $P_R=(1+\gamma_5)/2$ and $\sigma^{\mu \nu}
= {i\over 2} [\gamma_\mu,\gamma_\nu]$. As in \cite{GREUB}, we will set
$m_s=0$.

In this section, we present the details of the calculation of the
matrix element $\langle s\gamma|Q_2|b\rangle$ in the NDR scheme.  This
matrix element vanishes at the one-loop level. Therefore, in order
to obtain a non-vanishing contribution, one has to calculate two-loop
diagrams.  They are shown in Figs.~\ref{fig:M1}, \ref{fig:M2},
\ref{fig:M3}, and \ref{fig:M4}, where the wavy and dashed lines
represent the photon and gluon, respectively.  Following
\cite{GREUB}, we have divided the contributing diagrams into
four sets:
\begin{itemize}
\item In sets 1 and 2, the photon is emitted from an internal s-quark
  (set1) or an internal b-quark (set 2), which also exchanges a gluon
  with the charm quark.
\item Sets 3 and 4 are obtained from sets 1 and 2, respectively, by
emitting this time the photon from the charm quark propagator.
\item Thanks to QED gauge invariance, it is not necessary to consider
  diagrams with a real photon emission from external quark lines.
\end{itemize}

It is convenient to write the regularized contributions from each set
of the diagrams as follows:
\ba
M_1 &=& \left\{ {1\over 36\ep} \left( {m_b\over \kappa \mu}\right)^{-4\ep}
 + \widetilde{M}_1\right\} {\alpha_s\over \pi}\, C_F Q_d 
 \,\langle s\gamma|Q_7|b\rangle_{\rm tree},
\label{eq1}
\\
M_2 &=& \left\{ -{5\over 36\ep} \left( {m_b\over \kappa \mu}\right)^{-4\ep}
 + \widetilde{M}_2\right\} {\alpha_s\over \pi}\, C_F Q_d 
 \,\langle s\gamma|Q_7|b\rangle_{\rm tree},
\label{eq2}
\\
M_3 &=& \left\{ -{1\over 8\ep} \left( {m_b\over \kappa \mu}\right)^{-4\ep}
 + \widetilde{M}_3\right\} {\alpha_s\over \pi}\, C_F Q_u 
 \,\langle s\gamma|Q_7|b\rangle_{\rm tree},
\label{eq3}
\\
M_4 &=& \left\{ -{1\over 4\ep} \left( {m_b\over \kappa \mu}\right)^{-4\ep}
 + \widetilde{M}_4\right\} {\alpha_s\over \pi}\, C_F Q_u 
 \,\langle s\gamma|Q_7|b\rangle_{\rm tree}.
\label{eq4}
\ea
We work in $D=4-2\ep$ dimensions, with $\kappa^2 = 4\pi e^{-\gamma_E}$,
$Q_u = \f{2}{3}$, $Q_d=-\f{1}{3}$, $C_F=\f{4}{3}$ and $\gamma_E$ is
Euler's constant. The tree level
matrix element of the operator $Q_7$ is given by
\ba
 \,\langle s\gamma|Q_7|b\rangle_{\rm tree} = m_b(\mu_b){e\over 8\pi^2}
 \overline{u}(p') \epsla \qsla P_R u(p).
\ea

In order to make the two-loop matrix element finite, counterterms
have to be added.  These counterterms can easily be calculated by
using the operator renormalization constants that were needed in the
context of the calculation of the leading order anomalous dimension
matrix.  The complete counterterm is found to be \cite{GREUB}
\ba
M_2^{\rm count} = \left\{ {Q_d\over 6}  \left( {m_b\over \mu}
\right)^{-2\ep}  + {3Q_u\over 8} -{Q_d\over 18}\right\}
{\kappa^{4\ep} \over \ep} \; {\alpha_s \over \pi} \, C_F 
 \,\langle s\gamma|Q_7|b\rangle_{\rm tree}.
\label{eq6}
\ea
The above expression includes two-loop counterterms as well as
contributions from one-loop diagrams with one-loop counterterm
insertions.

Adding the contributions from (\ref{eq1})--(\ref{eq4}) and
(\ref{eq6}), we find the two-loop matrix element $\langle
s\gamma|Q_2|b\rangle$ in the NDR-scheme:
\ba
\langle s\gamma|Q_2|b\rangle =  
 \,\langle s\gamma|Q_7|b\rangle_{\rm tree} {\alpha_s\over 4\pi}
 \left( {416\over 81} \ln {m_b\over \mu} + r_2\right),
\ea
with 
\ba
\label{labeleqr2}
r_2 = -{16\over 9} (\widetilde{M}_1 + \widetilde{M}_2)
 + {32\over 9} (\widetilde{M}_3 + \widetilde{M}_4).
\ea
The rest of this section is devoted to a detailed presentation of the
calculation of the contributions $\widetilde{M}_i$ using the method of
asymptotic expansions. All the momentum integrals are performed in 
the Euclidean space.

\subsection{Diagram $M_1$}

\begin{figure}[htb]
\hspace*{60mm}
\psfig{figure=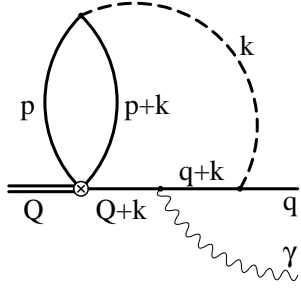,width=40mm}
\caption{Diagram $M_1$. Momentum assignments: $Q$, $q$, and $\gamma$ denote
respectively 
four-momenta of the $b$ and $s$ quarks, and of the photon; $p$ and $k$
are loop momenta.  Double lines denote a massive particle (the $b$
quark) and single lines are light particle propagators. The gluon is denoted
by a dashed line.} 
\label{fig:M1}
\end{figure}

\begin{figure}[htb]
\hspace*{60mm}
\psfig{figure=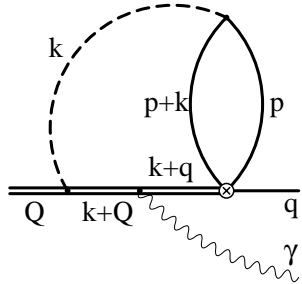,width=40mm}
\caption{Diagram $M_2$.}
\label{fig:M2}
\end{figure}

Of  the four sets of diagrams which we have to consider, $M_1$ and
$M_2$ shown in Fig.~\ref{fig:M1} and \ref{fig:M2} are relatively
simple. In $M_1$, the photon is emitted from a massless $s$-quark
propagator, and the two-loop integral factorizes into the $c$-quark
loop and the one-loop vertex integral.  There are four momentum regions
we have to consider.  Denoting the momenta in the $c$-quark and vertex
loops by $p$ and $k$, we have
\begin{enumerate}
\item $k\sim p \sim m_b$: the ``hard-hard'' region;
\item $k\sim m_b$ and $p \sim m_c$: the ``hard-soft'' region (which
 comes in two variants, depending on the routing of $k$ through the
 $c$-quark loop);
\item a ``collinear-soft'' region, in which $p\sim m_c$ and $k\cdot \gamma
 \sim m_b^2$ but $k\cdot q \sim m_c^2$.  
\end{enumerate}
Below, we describe the treatment of those regions in some detail.

\subsubsection{Hard contribution: Taylor expansion in $m_c$}

\begin{figure}[htb]
\hspace*{60mm}
\psfig{figure=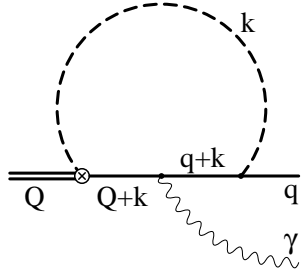,width=40mm}
\caption{One-loop diagram (the $k$ propagator can have fractional
power) corresponding to eq.~(\protect\ref{eqk1}).}
\label{fig:oneVert}
\end{figure}

Since all scalar products are now large compared with $m_c^2$, we can
expand the $c$-quark propagators in $m_c$.   This leads to a massless
loop integration, which simply modifies the power of the momentum
($k$) in the gluon propagator (see Fig.~\ref{fig:oneVert}). 
After this first integration, the integrand has the following form:
\ba
\int 
{ \dd ^D k\; \mbox{(scalar products involving $k$)}
\over 
(k^2)^{a_1+\vep} (k^2+2k\cdot Q+Q^2)^{a_2} (k^2+2k \cdot q)^{a_3}
}.
\label{eqk1}
\ea
We combine propagators 1 and 3 using the Feynman parameter $x$, and
then combine the result ($\cdot (1-y)$) with propagator 2 ($\cdot y$).
We shift the variable $k\to K-x(1-y)q-yQ$, and the denominator becomes
a power of $k^2+y(1-y)(x-1-i0)$.  Since $x-1<0$, this diagram has an
imaginary part; its sign is determined by assigning a small negative
imaginary mass to the massless lines.  After the momentum shift, 
we have additional factors $x^m$ and $y^n$
in the numerator, and denominator has the power $d+\vep$.  The result
is
\ba
&&
\left[ \exp(-i\pi)\right]^{-2\vep}
(-1)^d{\Gamma(d-2+2\vep)\over \Gamma(d+\vep)}
{B(m+a_3,a_1+2-d-\vep)\over B(a_3,a_1+\vep)B(a_2,a_1+a_3+\vep)}
\nonumber \\
&&
\times B(n+a_2+2-d-2\vep, a_1+a_3+2-d-\vep),
\ea
where $B$ and $\Gamma$ are the standard Euler functions Beta and
$\Gamma$. 

\subsubsection{Hard-soft contributions from $k\sim m_b$, $p\sim m_c$}
\label{sec:mun1cRL}

\begin{figure}[htb]
\hspace*{60mm}
\psfig{figure=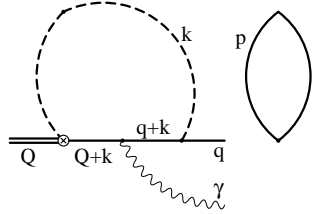,width=40mm}
\caption{Subgraph corresponding to one of the two hard-soft
             contributions.} 
\label{fig:Munich1cR}
\end{figure}

This region can be handled as in the ``large momentum expansion''
\cite{Chetyrkin91,Tkachev:1994gz,Smirnov:1995tg}.  In
Fig.~\ref{fig:Munich1cR}, we see one of the two configurations to be
computed.  We first integrate over $p$, and then evaluate the one-loop
vertex.  An
analogous procedure is executed for the other (left) $c$-quark line.

\subsubsection{Collinear region}
Naively, one might think that we should consider the region $k\sim p\sim
m_c$, in which $(Q+k)^2\simeq Q^2$, $(q+k)^2\simeq 2k\cdot q$.  However,
this does not lead to a consistent power counting, since we have to shift
the momentum $k$ by an amount proportional to $q$, which is $\sim m_b$.  

Instead, we have to consider the collinear configurations
\cite{Smirnov:1997gx}.   We define
\ba
k\cdot q = Q_0 k_-,&&   k\cdot\gamma = Q_0 k_+, 
\quad k^2 = k_+ k_- -k_\perp^2,
\quad Q^2 = Q_0^2, \quad k\cdot Q = Q_0(k_++k_-).
\nonumber \\
&&
\ea
The relevant contribution is $k_+\sim Q_0$, $k_-\sim 
{m_c^2\over Q_0}$.  Now $(Q+k)^2 \simeq Q^2
 + 2k\cdot\gamma$, while $(k+q)^2=
k^2+2k\cdot q$ is already homogeneously of order $m_c^2$ and
cannot be 
expanded.  First, we integrate over $p$; for this we combine the two
$c$ propagators (we assume here, for generality, that their powers are
$a_1$ and $a_2$) using $x$, shift $p\to P-xk$, and obtain ($m$ and $d$
are exponents arising from the presence of $p$ in the numerator)
\ba
{{\rm polynomial}(p)\over 
\left[ (p+k)^2+m_c^2\right] ^{a_1}
\left[ p^2+m_c^2\right] ^{a_2}}
&\to &
{\Gamma(d-2+\vep)\over B(a_1,a_2)\Gamma(d)}
{x^{m+a_1+1-d-\vep} (1-x)^{a_2+1-d-\vep}
\over (k^2+m_x^2)^{d-2+\vep}},
\nonumber \\
 m_x^2 &\equiv& {m_c^2\over x(1-x)}.
\ea 
For the $k$ integration, the integrand has the form
\ba
\int {\dd^D k \;\; {\rm polynomial}(k)\over 
(k^2)^{a_3} (k^2+m_x^2)^{d-2+\vep}
(k^2+2k\cdot q)^{a_4}  (Q^2+2\gamma\cdot k)^{a_5}
}.
\ea
Now, we combine the first two propagators multiplying them by $1-z$
and $z$, respectively.  We multiply the result by $y$ and combine it
with the ``$a_4$'' propagator multiplied by $1-y$.  Finally, we use
the parameter $u$ to include the last propagator
($i=a_3+a_4+d-2+\vep$): 
\ba
\lefteqn{
 {1\over [k^2+yzm_x^2+2(1-y)k\cdot q]^i\;
(Q^2+2k\cdot \gamma)^{a_5} }}
\nonumber \\&&
= {1\over B(i,a_5)}\int_0^\infty \dd u\; {u^{a_5-1} \over
[k^2+yzm_x^2+2(1-y)k\cdot q+2uk\cdot\gamma+uQ^2]^{i+a_5}}.
\ea
Now, we shift $k\to K-(1-y)q-u\gamma$ and, using $q\cdot\gamma = Q^2/2$,
find that
the denominator simplifies to become $[ K^2 + y(zm_x^2+uQ^2)]$.
Because of the presence of $K$ in the numerator, we generate
additional powers of Feynman parameters. After integrating over
$K$, we integrate over $u$,
\ba
\int_0^\infty \dd u \; {u^{n-1} \over 
(zm^2 +uQ^2)^m } 
= 
{B(n,m-n)\over (Q^2)^n (zm^2)^{m-n}}.
\ea
We see that the dependence on $Q$ separates (so the integral really
depends only on a single scale).  Also, only $m_c$ appears in fractional
power, so that there is no imaginary part in this integral, even
though $Q^2<0$ (after the Wick rotation).

\subsection{$M_2$}
In this diagram, we have again four momentum regions, of which the first
three are similar to those discussed for $M_1$.  The fourth one is
different: it is a ``soft-soft'' contribution, with $k\sim p \sim
m_c$.  

For the hard-hard contribution, we first integrate over $p$, just like
in $M_1$.   After that, the integrand takes on the following form:
\ba
\int 
{ \dd ^D k\; \mbox{(scalar products involving $k$)}
\over 
(k^2)^{a_1+\vep} (k^2+2k\cdot Q)^{a_2} (k^2+2k\cdot q-Q^2)^{a_3}
}.
\ea
We combine propagators 2 and 3 using a Feynman parameter $x$, and
then combine the result ($\cdot y$) with the propagator 1 ($\cdot
(1-y)$).  We shift the variable $k\to K-y((1-x)Q+xq)$, and the denominator
becomes a power of $k^2+y[y+x(1-y)]$.    After  shifting the momentum and
simplifying the numerator, additional factors $x^m$ and $y^n$ appear
in the numerator, and the denominator has the power $d+2\vep$.  After the
$k$ integration, we find  that the $x$ and $y$ integrations do not
separate, and we first integrate over $x$ using
eq.~(\ref{eq:xInteg}), to be discussed below.  The remaining $y$
integration is a Beta function.  

The hard-soft contributions are similar to those in $M_1$ and lead to
rather trivial products of one-loop integrals.  

On the other hand, the soft-soft contribution is less standard.  We
have $k\sim m_c$ and the two $b$-quark propagators can be expanded.
In the lowest order we have 
\ba
{1\over k^2 + 2k\cdot Q} &\to& {1\over 2k\cdot Q},
\nonumber \\
{1\over k^2 + 2k\cdot q -  Q^2} &\to& -{1\over Q^2}.
\ea
Since the second propagator becomes independent of $k$, the whole
two-loop diagram becomes equivalent to a simpler, two-point function.
The resulting integrals are similar, but not the same, as those
considered in the eikonal expansion study \cite{CzSmir}. 
Details of their evaluation have been recently
described in a completely different context of bound-state
calculations \cite{Czarnecki:2000fv}.  In that work, exactly such type
of integrals appeared when energies of bound-states consisting of two
particles with very different masses were expressed  as expansions in
the ratio of those masses.

\subsection{$M_3$}
\begin{figure}[htb]
\hspace*{30mm}
\begin{tabular}{c@{\hspace*{15mm}}c}
\psfig{figure=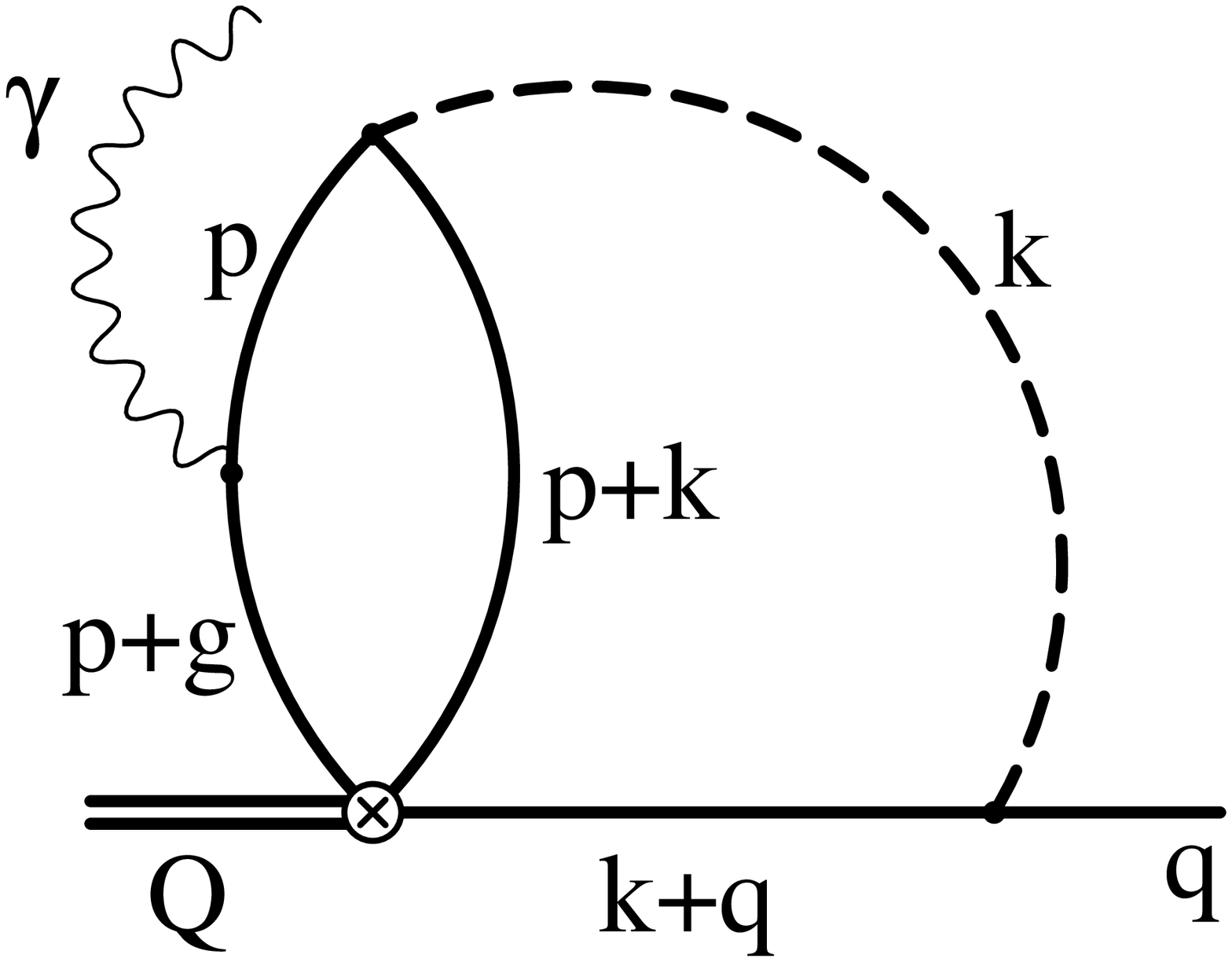,width=40mm}
&
\psfig{figure=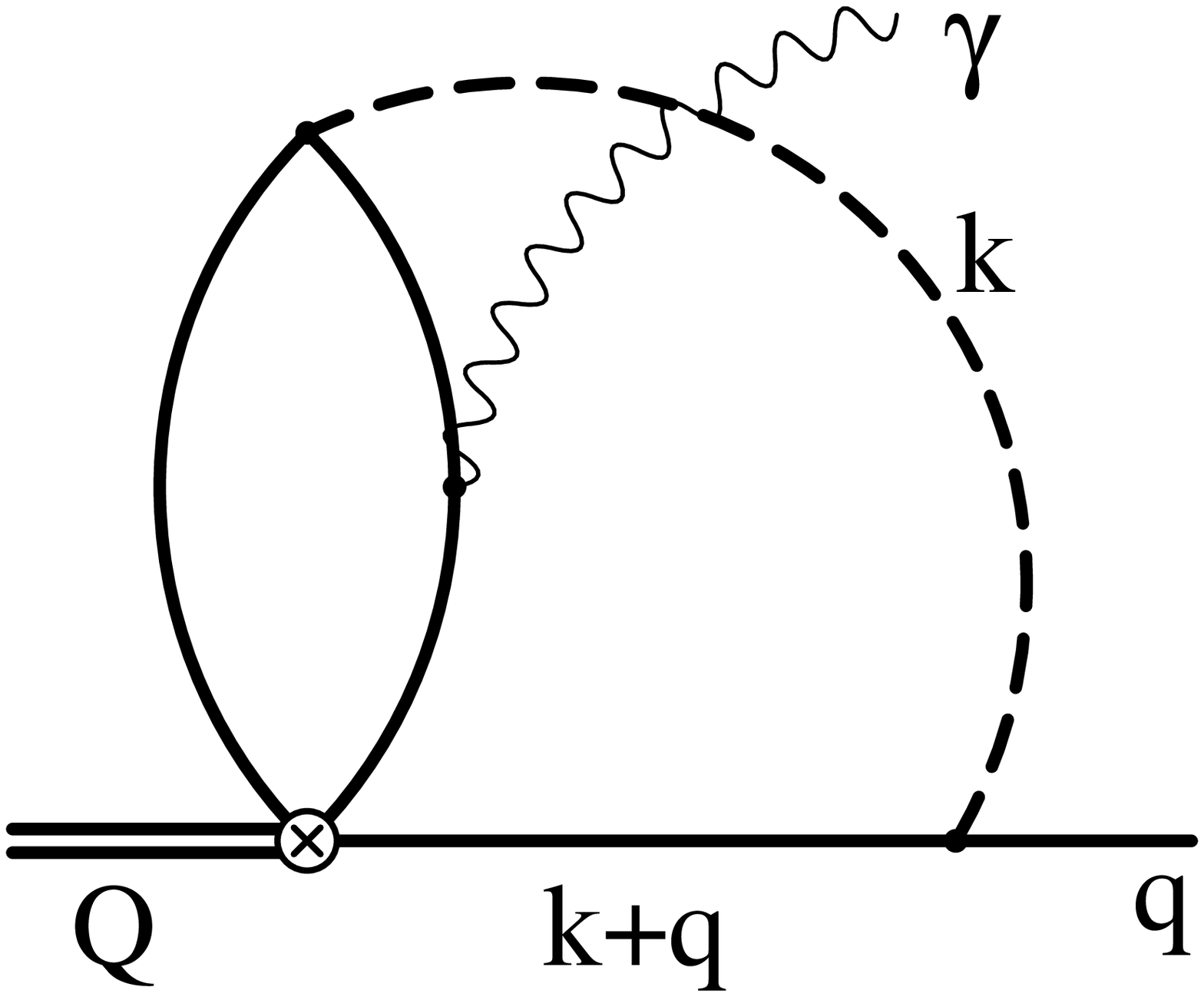,width=40mm}\\[1mm]
(a) & (b)
\end{tabular}
\caption{Diagrams contributing to $M_3$, with photon emission from two
$c$-quark lines.}
\label{fig:M3}
\end{figure}
This diagram, shown in Fig.~\ref{fig:M3}, is the first of the two
non-trivial two-loop vertex diagrams we have to consider.  There are
now five momentum regions to be considered:
\begin{enumerate}
\item Hard-hard, $k\sim p \sim m_b$, similar to those of $M_{1,2}$. 
\item Hard-soft, $k\sim m_b$, $p\sim m_c$, which now enters only once,
when hard momentum flows through the same $c$-quark line from which
the photon is emitted.
\item Collinear-collinear, with $k$ and $p$ having their only
large components ($\order{m_b}$) aligned with the $s$-quark momentum
$q$ (and with $k^2\sim p^2 \sim m_c^2$). 
\item Collinear-collinear, but with the alignment with the photon
momentum $\gamma$. 
\item Hard-collinear, where $k\sim m_b$ and $p$ is aligned parallel
with $\gamma$. 
\end{enumerate}

The first two regions are treated in an analogous manner as was
described for $M_{1,2}$.  In the hard-hard contribution, after
expansion in $m_c$, the integrand is symmetrical (apart from different
powers of propagators) under replacement $p\leftrightarrow k$,
$\gamma\leftrightarrow q$.  Hence, we can first integrate over $k$,
and then over $p$.  Both integrations are very similar to those in
$M_{1,2}$.  In the second region, we use the ``large momentum
expansion'' mentioned in Section~\ref{sec:mun1cRL}.

The collinear-collinear regions require more attention
\cite{Smirnov:1999bz}.  After expansion of the integrands in the
available small quantities, it turns out that the integrals over
Feynman parameters have singularities (like $\int_0 \dd x/x$) which
are not regularized by our dimensional regulator.  It is necessary to
introduce additional, analytical regularization on the heavy quark
line.  Similar phenomena have been observed before (see
e.g.~\cite{Smirnov:1999bz} and references therein).  Nevertheless, all
integrals over the Feynman parameters can be evaluated analytically
without particular difficulties.  Dependence on the analytical
regulator cancels in the sum of the two doubly-collinear
contributions. 

\subsection{$M_4$}
\begin{figure}[htb]
\hspace*{30mm}
\begin{tabular}{c@{\hspace*{15mm}}c}
\psfig{figure=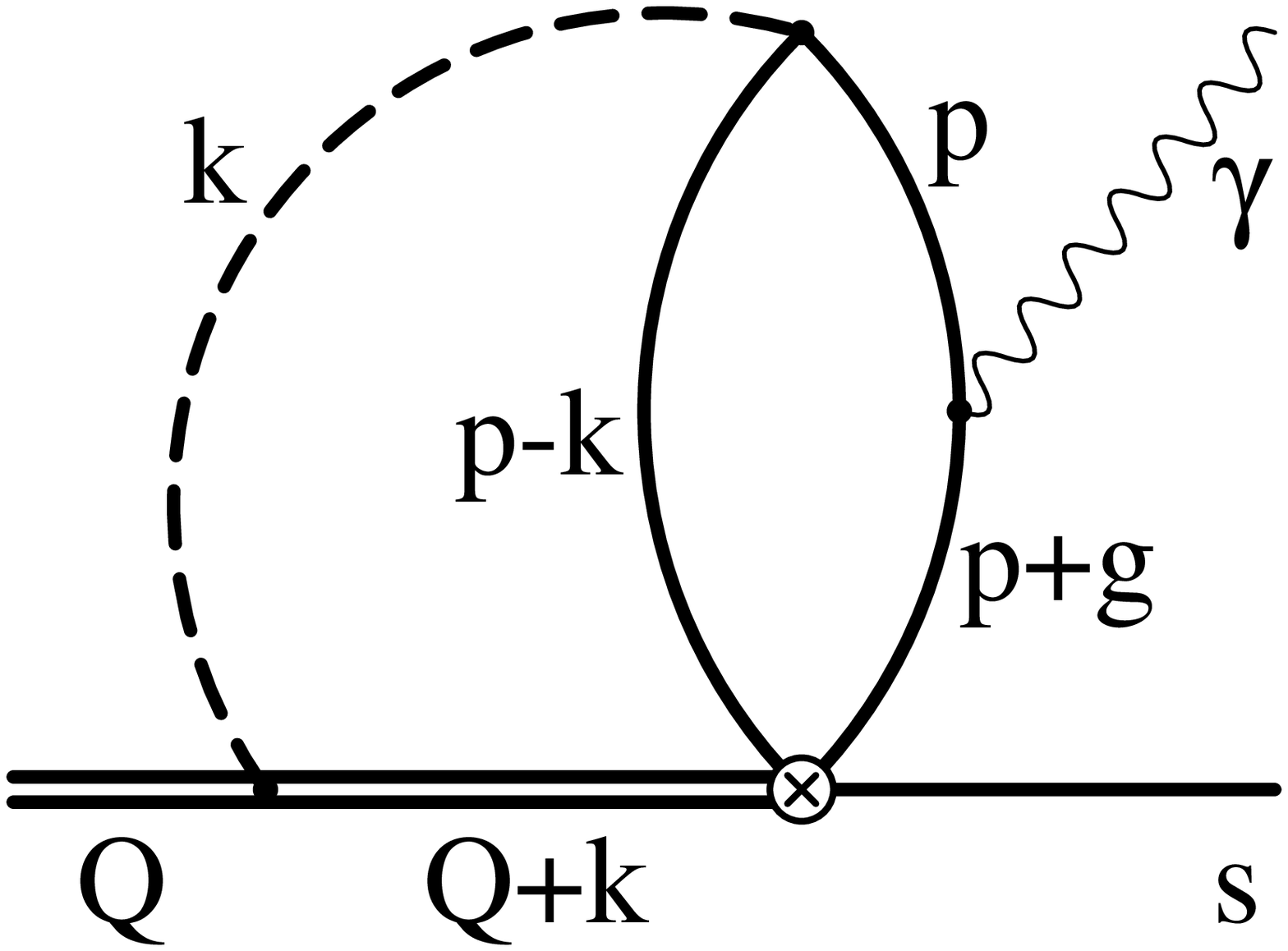,width=40mm} &
\psfig{figure=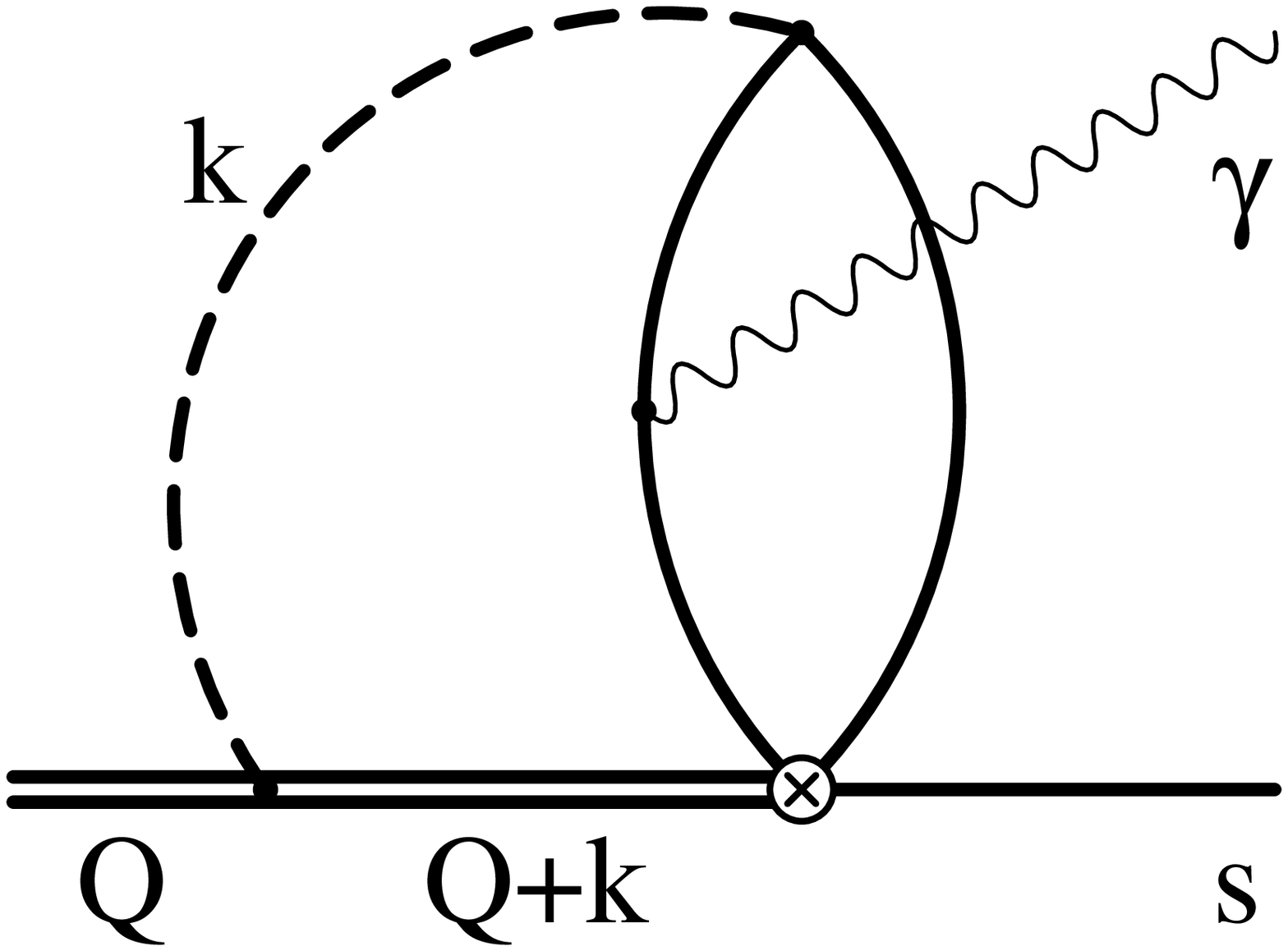,width=40mm}\\[1mm]
(a) & (b)
\end{tabular}
\caption{Diagram $M_4$}
\label{fig:M4}
\end{figure}

The most complicated diagram is $M_4$, depicted in Fig.~\ref{fig:M4}.  
There are six momentum regions:
\begin{enumerate}
\item Hard-hard.
\item Hard-soft.
\item Collinear-collinear with alignment along $\gamma$.  There is no
contribution with alignment along $q$ here, because of the different
structure of the internal quark propagator, which now contains the large
$b$-quark mass.
\item Hard-collinear, as in $M_3$. 
\item Ultrasoft-collinear \cite{Smirnov:1999bz}, 
with $k\sim m_c^2/m_b$ and $p$ aligned with $\gamma$. 
\item Soft-soft.
\end{enumerate}

We give a more detailed description of the hard-hard contribution,
because the integrals resulting here are rather complicated.  First, we
consider the scalar integral with all propagators in the first power,
\ba
J_3 =\int  { \dd^D k \over k^2 (k^2+2Q\cdot k)}
 \int { \dd^D p \over p^2 (p-k)^2 (p+\gamma)^2}.
\ea
We first evaluate the massless ($c$ quark) loop (integrals over all
Feynman parameters run from 0 to 1):
\ba
\int { \dd^D p \over p^2 (p-k)^2 (p+\gamma)^2} &=& 
\int { \dd^D p \over p^2 (p^2-2p\cdot k+k^2) (p^2+2p\cdot \gamma)} 
\nonumber \\
&=& \int 2(1-y)\dd y\; \dd x\; 
\int {\dd^D p \over  [p^2-2yp\cdot k + 2x(1-y)p\cdot \gamma + yk^2]^3 }
\nonumber \\
&=& \Gamma(1+\vep)\, \pi^{2-\vep} \int {y^{-1-\vep}(1-y)^{-\vep}\dd y\; \dd x  \over
 (k^2 +2xk\cdot \gamma)^{1+\vep} } 
\nonumber \\
&=& \Gamma(1+\vep)B(-\vep,1-\vep)\,  \pi^{2-\vep}
\int { \dd x  \over (k^2 +2xk\cdot \gamma)^{1+\vep} } .
\label{eq:intP}
\ea
In the process of the integration, we made a shift $p\to p+yk - x(1-y)\gamma$.
Next, we integrate over $k$:
\ba
\lefteqn{\int \dd x
\int  { \dd^D k \over k^2 (k^2+2Q\cdot k) (k^2 +2xk\cdot \gamma)^{1+\vep}  }}
\nonumber \\
&=& (1+\vep)
 \int \dd x\; \dd u\; (1-u)^{\vep}\int  { \dd^D k \over k^2 
      [k^2+2uk\cdot Q +2(1-u)xk\cdot \gamma]^{2+\vep}  } 
\nonumber \\
&=&  (2+\vep)(1+\vep) \int \dd z\; z^{1+\vep}\dd x\; \dd u\; (1-u)^{\vep}\int  
 { \dd^D k \over [k^2+z^2u^2+z^2u(1-u)x]^{3+\vep}  } 
\nonumber \\
&=& {\Gamma(1+2\vep)\over \Gamma(1+\vep)}\, \pi^{2-\vep}
\int \dd z\;\dd x\; \dd u\; {z^{-1-3\vep} u^{-1-2\vep}(1-u)^{\vep} \over 
 [u+(1-u)x]^{1+2\vep}  } 
\nonumber \\
&=& {\Gamma(1+2\vep)\over 6\vep^2\Gamma(1+\vep)}\, \pi^{2-\vep}
\int  \dd u\;  u^{-1-2\vep}(1-u)^{-1+\vep}(1-u^{-2\vep})
\nonumber \\
&=& {\Gamma(1+2\vep)\over 6\vep^2\Gamma(1+\vep)} \,\pi^{2-\vep}
\left[ B(-2\vep,\vep)-B(-4\vep,\vep)\right].
\label{eq:intK}
\ea
We made a shift $k\to k-zuQ -z(1-u)x\gamma$ and used $Q^2= 2Q\cdot\gamma = -1$. 
Multiplying the result (\ref{eq:intK}) with the coefficient
$\Gamma(1+\vep)B(-\vep,1-\vep)$ from (\ref{eq:intP}) we find
\ba
J_3 &=&  
{\Gamma(1+2\vep)\over 6\vep^2} B(-\vep,1-\vep)
\left[ B(-2\vep,\vep)-B(-4\vep,\vep)\right] 
\nonumber \\[1mm]
&=&
\Gamma(1+\vep)^2 \left({1\over 24\vep^4} + {\pi^2\over 18\vep^2}
+{7\over 6\vep} \zeta_3 + {29\pi^4 \over 360}\right) + \order{\vep}.
\ea

Now we consider the general case, in which momenta $p$ and $k$
can be present in the numerator,
\ba
\int 
{ \dd ^D k\; \dd ^D p\; \mbox{(scalar products involving $k$ and $p$)}
\over 
{(k^2)}^{a_1} {(k^2+2k\cdot Q)}^{a_2} {(p-k)}^{2a_3} {(p^2)}^{a_4}
{(p^2+2p\cdot \gamma)}^{a_5}
}.
\ea
We first combine the propagators 5 ($\cdot x$) and 4 ($\cdot (1-x)$),
and then the result ($\cdot (1-y)$) with 3 ($\cdot y$).  We shift the
integration momentum $p = K-x(1-y)\gamma+yk$ and average over $K$.
This again results in extra powers $y^n$ and $x^m$, while canceling
powers of $K^2$ changes the power of the denominator.  After the
integration over $K$, the denominator becomes
$[y(1-y)(k^2+2xk\cdot\gamma)]^{d-2+\vep}$.  We see that the dependence
on $y$ factorizes and we can integrate over this variable.  As a
result, we find an expression of the form
\ba
{
 \Gamma(a_3+a_4+a_5)
 \Gamma(d-2+\vep)
 B(n+a_3-d+2-\vep, a_4+a_5-d+2-\vep)
\over
 \Gamma(a_3) \Gamma(a_4) \Gamma(a_5) \Gamma(d)
}
&&
\label{eq:factorP1}
\\
\times \int \dd x\; x^{a_5+m-1}(1-x)^{a_4-1}
\rule{30mm}{0mm}
&&
\label{eq:factorP2}
 \\
\times
\int 
{ \dd ^D k\; \mbox{(scalar products involving $k$)}
\over 
{(k^2)}^{a_1} {(k^2+2k\cdot Q)}^{a_2}{(k^2+2xk\cdot\gamma)}^{d-2+\vep}
}.\rule{2mm}{0mm}&&
\ea
We now repeat a similar procedure with the variable $p$.  We combine
the propagator 2 ($\cdot u$) with $(k^2+2xk\cdot\gamma)$ ($\cdot
(1-u)$), and then the result ($\cdot z$) with the propagator 1 ($\cdot
(1-z)$).  We change the momentum variable $k=K-zuQ-zx(1-u)\gamma$,
average over $K$ and simplify powers of $K^2$ (as a result, the power
of the denominator changes from $a_1+a_2+d-2$ to some $d_1$, and we
also get extra powers $z^e$ and $u^w$).  After integrating over $K$ we
get (including the factors (\ref{eq:factorP1},\ref{eq:factorP2}))
\ba
&&{
 \Gamma(a_3+a_4+a_5)
 \Gamma(d-2+\vep)
 B(n+a_3-d+2-\vep, a_4+a_5-d+2-\vep)
\over
 \Gamma(a_3) \Gamma(a_4) \Gamma(a_5) \Gamma(d)
}
\nonumber \\
&\times &
{\Gamma(d_1-2+2\vep)
\Gamma(a_1+a_2+d-2+\vep)
B(a_1,a_2+d+2+e-2d_1-3\vep)
\over
\Gamma(d_1+\vep)
\Gamma(a_1)
\Gamma(a_2)
\Gamma(d-2+\vep)
}
\nonumber
\\
& \times & 
\int \dd x\; \dd u\; 
x^{a_5+m-1}(1-x)^{a_4-1}
u^{w+a_2+1-d_1-2\vep} 
(1-u)^{d-3+\vep }
\left[ u+x(1-u) \right]^{2-d_1-2\vep}.
\rule{10mm}{0mm}
\ea
First we integrate over $x$, using
\ba
\int_0^1 \dd x\; x^n \left[ u+x(1-u) \right]^\mu
 = {u^{\mu+n+1}\over (1-u)^{n+1} }
\sum_{j=0}^n
(-1)^j {n \choose j}
{u^{j-1-n-\mu}-1\over \mu+n+1-j}.
\label{eq:xInteg}
\ea
This integration is possible because the powers of $x$ and $(1-x)$ in
(\ref{eq:factorP2}) are non-negative integer.  
The $u$ integration gives a simple Beta function.

The techniques described above are sufficient to compute all the
remaining contributions.  Again, the dimensional regularization alone
is insufficient to evaluate the doubly-collinear contribution.  We
introduce an analytical regulator on the heavy quark mass.  The
resulting singularities vanish when we add the soft-soft
contribution.

\subsection{Results}

The final results for $\widetilde{M}_i$ are listed below, with
$z=m_c^2/m_b^2$ and $L = \ln z$. 

\begin{eqnarray}
\widetilde M_1 &=& 
    {37 \over 216} 
       + \left(  - {5 \over 2} - L \right)z
       +  \left(  - {5 \over 2} + \pi^2 + L - L^2 \right)z^2
\nonumber \\ 
&& + \left(  - {17 \over 27} - {2 \over 3} \pi^2 - {10 \over
       9} L + {2 \over 3} L^2 \right) z^3
\nonumber 
\\
&&
       +  \left(  - {11 \over 12} + L \right)z^4
       +  \left(  - {1 \over 30} + {2 \over 3} L \right)z^5
       +  \left( {67 \over 270} + {7 \over 9} L \right)z^6
\nonumber \\
&&
 + i\pi \left[ {1\over 18}
 -z  + (  1- 2L  )z^2 + \left(-{10\over 9} + {4\over 3}L\right)z^3
 +z^4 + {2\over 3}z^5 + {7\over 9}z^6 \right],
\label{eqM1}
\\[2mm]
\widetilde M_2& =&   
  {13 \over 216}      
  +   \left(  - {1 \over 2} + {1 \over 6} \pi^2 \right)z
  -  {2 \over 3} \pi^2  z^{{3 \over 2}}   
  +  \left( 3 - 3 L + {1 \over 2} L^2 \right)z^2
\nonumber \\ &&
  + \left(  - {157 \over 108} + {5 \over 9} \pi^2 
           + {1 \over 18} L + {4 \over 3} L^2 \right)z^3  
  + \left(  - {4679 \over 900} + {2 \over 3} \pi^2
             + {107 \over 30} L + 2 L^2 \right)z^4
\nonumber \\ &&
  + \left(  - {26185 \over 2352} + {5 \over 3} \pi^2 
       + {277 \over 28} L + 5 L^2 \right)z^5
\nonumber \\  &&
  + \left(  - {2831737 \over 97200} + {14 \over 3} \pi^2
        + {16177 \over 540} L + 14 L^2 \right)z^6,   
\\[2mm]
\widetilde M_3 &=& 
 - {15 \over 16} 
 + \left( {3 \over 2} - {1 \over 2} \pi^2 L - {1 \over 4} \pi^2 
    - 2 \zeta_3
      + 2 L + {1 \over 4} L^2 + {1 \over 6} L^3 \right) z
\nonumber \\ &&
 +   \left( {5 \over 4} - {1 \over 2} \pi^2 L
        + {1 \over 2} \pi^2 - 2 \zeta_3 + {1 \over 2} L 
        - {1 \over 2} L^2 + {1 \over 6}   L^3 \right)  z^2 
\nonumber \\ &&
 +   \left(  - {17 \over 12} + L \right) z^3
 +    \left(  - {7 \over 216} + {5 \over 9} L \right)z^4
\nonumber \\ &&
 +   \left( {1183 \over 4320} + {49 \over 72} L \right)z^5 
 +    \left( {8783 \over 12000} + {231 \over 200} L \right)z^6
\nonumber \\ &&
+i\pi\left[ - {1 \over 4} 
+ \left( {1 \over 2} L + {1 \over 2} L^2 + 2 
                     - {1 \over 6} \pi^2 \right)  z
 +  \left(  - L + {1 \over 2} L^2 + {1 \over 2} 
                  - {1 \over 6}\pi^2 \right) z^2
\right.
\nonumber \\
&& \left.\qquad
 +  z^3  +  {5 \over 9} z^4 
 +   {49 \over 72} z^5 
 + {231 \over 200} z^6
 \right] ,
%
\\[2mm]
\widetilde M_4 &=& 
 - {7 \over 8} 
+ \left( 1 - {1 \over 4} \pi^2 L - {1 \over 12} \pi^2 
      - \zeta_3 + {1 \over 2} L - {1 \over 12} L^3 \right)z
\nonumber \\ &&
+ \left( {1 \over 2} + {1 \over 6} \pi^2 + 2 \zeta_3 
     - {1 \over 2} L + {1 \over 4} L^2 - {1 \over 12} L^3 \right)z^2
\nonumber \\ &&
+ \left( {1 \over 4} - {1 \over 4} \pi^2 + L
                         - {3 \over 4} L^2 \right)z^3
\nonumber \\ &&
+ \left( {58 \over 27} - {5 \over 18} \pi^2 - {41 \over 72} L
                                   - {5 \over 6} L^2 \right)z^4
\nonumber \\ &&
+ \left( {6547 \over 1728} - {35 \over 72} \pi^2
            - {283 \over 144} L - {35 \over 24} L^2 \right)z^5
\nonumber \\ &&
+ \left( {185837 \over 24000} - {21 \over 20} \pi^2
            - {4227 \over 800} L - {63 \over 20} L^2 \right)z^6.
\label{eqM4}
\end{eqnarray}

For the imaginary part (and the leading power of $L$ in the real part)
it is possible to guess the form of the higher order terms.  For
example, for $\widetilde M_1$, beginning with $z^4$ we have
\ba
\mbox{Higher orders in Im}\,\widetilde M_1 = 8z^3\sum_{n=1}^\infty 
 { (2n+1)!\over n\;(n+1)!\; (n+3)! } z^n.
\label{im1}
\ea
If $m_c=m_b/2$, we expect the imaginary part to vanish.  Indeed, this
corresponds to $z=1/4$, at which point (\ref{im1}) gives ${43\over
288}-{5\over 24}\ln 2$, which exactly cancels the contribution of the
first 4 terms in the imaginary part of $\widetilde{M}_1$ given in 
(\ref{eqM1}). 
The results for the leading terms and the powers $z,\,z^{3/2},\,z^2$ 
and $z^3$
in (\ref{eqM1})--(\ref{eqM4}) agree with the corresponding terms in
(2.25)--(2.28) of Greub, Hurth and Wyler \cite{GREUB}. The
contributions $z^4,\, z^5$ and $z^6$ are new.\\
Inserting (\ref{eqM1})--(\ref{eqM4}) into (\ref{labeleqr2}) we find

\[
r_2 = {\rm Re}\, r_2 + i {\rm Im}\, r_2
\]
with \newpage
\begin{eqnarray}
{\rm Re}\, r_2 &=& 
   - {1666 \over 243}
   + {32 \over 27} \pi^2 z^{3/2} 
\nonumber\\
\nonumber
 &&+ \left( {128 \over 9}-{8 \over 3} \pi^2 L-{40 \over 27} \pi^2-
     {32 \over 3} \zeta_3+{32 \over 3} L+{8 \over 9} L^2+ 
     {8 \over 27} L^3 \right) z
\\\nonumber
 &&+ \left( {16 \over 3}-{16 \over 9} \pi^2 L+{16 \over 27} \pi^2+
     {32 \over 9} L+{8 \over 27} L^3 \right) z^2 
\\\nonumber
 &&+ \left( -{4 \over 9}-{56 \over 81} \pi^2+{728 \over 81} L-
     {56 \over 9} L^2 \right) z^3 
\\\nonumber
 &&+ \left( {111748 \over 6075}-{176 \over 81} \pi^2-{3308 \over
     405} L-{176 \over 27} L^2 \right) z^4 
\\\nonumber
 &&+ \left( {816731 \over 23814}-{380 \over 81} \pi^2-
     {13234 \over 567} L-{380 \over 27} L^2 \right) z^5 
\\
 &&+ \left( {44551813 \over 546750}-{1624 \over 135} \pi^2 -
     {421121 \over 6075} L-{1624 \over 45} L^2\right) z^6,
\\\nonumber
{\rm Im}\, r_2 &=& \pi \left[ - {80 \over 81} 
   + \left( {16 \over 9} L + {16 \over 9} L^2 + {80 \over 9}  - 
     {16 \over 27} \pi^2 \right) z
   + \left( {16 \over 9} L^2 - {16 \over 27} \pi^2 \right) z^2\right.\\
 &&\left.+ \left(  - {64 \over 27} L + {448 \over 81} \right) z^3
   +  {16 \over 81} z^4  + {100 \over 81} z^5 + {5516 \over
   2025} z^6 \right] 
\end{eqnarray}
confirming the results (2.37) and (2.38) of \cite{GREUB}, and
generalizing them to include terms ${\cal O}(z^4),\,{\cal O}(z^5)$ and
to ${\cal O}(z^6)$. The contributions of different terms $z^n$ for
$m_c/m_b=0.29$ and $m_c/m_b=0.22$ are presented in table \ref{tabel1}.
The term $z^{3/2}$ has been added to the
term $z^2$ in this table.

\begin{table}[htb]
\begin{center}
\begin{tabular}{|c|r|r||r|r|}
\hline
  &\multicolumn{2}{|c||}{${\rm Re}\, r_2$} 
  &\multicolumn{2}{|c|}{${\rm Im}\, r_2$}
\\
\hline
$n$ & $\frac{m_c}{m_b} = 0.29$ & $\frac{m_c}{m_b} = 0.22$
    & $\frac{m_c}{m_b} = 0.29$ & $\frac{m_c}{m_b} = 0.22$
\\
\hline
0 &$ -6.8559671$ &$ -6.8559671$ &$ -3.1028076$ &$ -3.1028076$ \\
1 &$  2.2271232$ &$  1.6504869$ &$  2.5193530$ &$  2.1225820$ \\
2 &$  0.5775520$ &$  0.2307186$ &$  0.1121646$ &$  0.0769360$ \\
3 &$ -0.0402440$ &$ -0.0103793$ &$  0.0213018$ &$  0.0045268$ \\
4 &$ -0.0011397$ &$ -0.0002090$ &$  0.0000310$ &$  0.0000034$ \\
5 &$ -0.0001703$ &$ -0.0000187$ &$  0.0000163$ &$  0.0000010$ \\
6 &$ -0.0000307$ &$ -0.0000020$ &$  0.0000030$ &$  0.0000001$ \\
\hline
\end{tabular}
\caption{ The numerical value of {\it Coefficient} $\cdot z^n$ for the
  real and imaginary part of
  $r_2$ are given up to $z^6$ using two different values of
  $m_c/m_b$. The term $z^{3/2}$ in Re$\,r_2$ was included in the
  term proportional to $z^2$.\label{tabel1}}
\end{center}
\end{table}

We observe that the terms ${\cal O}(z^n)$ with $n\ge4$ are negligible.
The final result for $r_2$ is given by:
\begin{equation}
{\rm Re}\, r_2= \left\{\begin{array}{ll} -4.093\quad & m_c/m_b = 0.29\\
                                        -4.985 & m_c/m_b = 0.22
                      \end{array}\right.\;,
\qquad
{\rm Im}\, r_2 =\left\{\begin{array}{ll} -0.450\quad & m_c/m_b = 0.29\\
                                        -0.899 & m_c/m_b = 0.22
                      \end{array}\right..
\end{equation}

The strong dependence of $r_2$ on $m_c/m_b$ has been pointed out in
\cite{GAMMIS}. With decreasing $z$ the enhancement of $B\rightarrow X_s
\gamma$ by QCD corrections becomes stronger.

\section{Conclusion}
In the present paper, we have calculated the important two-loop matrix
element $\langle s\gamma |Q_2| b\rangle$ contributing to the decay
$B\rightarrow X_s \gamma$ at the NLO level.  Our result for $\langle
s\gamma |Q_2| b\rangle$ agrees with the one presented in \cite{GREUB}
and used in the literature by many authors.  The additional terms in
the expansion in $z=m_c^2/m_b^2$, that is $\ord(z^4)$ and higher, turn
out to amount to at most $0.05 \%$ and are negligible. As we have used
a completely different method from the one used in \cite{GREUB}, the
confirmation of the result of these authors is very gratifying.

\section*{Acknowledgments}
This work was supported by Deutscher Akademischer Austauschdienst
(DAAD), by the Natural Sciences and Engineering Research Council
(NSERC), and by the German Bundesministerium f{\"u}r Bildung und
Forschung under contract 05HT9WOA0. M.M. was supported by the Polish
Committee for Scientific Research under grant 2~P03B~121~20.

\vfill\eject


\end{document}